\documentclass[aps,amssymb,amsmath,pre,floatfix]{revtex4}
\usepackage{graphics}
\usepackage{psfrag}
\usepackage{epsfig}
\usepackage{subfigure}
\begin{document}
\bibliographystyle{prsty}

\title{Momentum Transport in Granular Flows}
\author{Gregg Lois$^{(1)}$}
\author{Ana\"el Lema\^{\i}tre$^{(2)}$}
\author{Jean M. Carlson$^{(1)}$}
\affiliation{
$^{(1)}$ Department of Physics, University of California, Santa Barbara, California 93106, U.S.A.}
\affiliation{$^{(2)}$ Institut Navier-- LMSGC, 2 all\'ee K\'epler, 77420 Champs-sur-Marne, France}

\date{\today}
\begin{abstract}
We investigate the error induced by only considering binary collisions in the momentum transport of hard-sphere granular materials, as is done in kinetic theories.  In this process, we first present a general microscopic derivation of the momentum transport equation and compare it to the kinetic theory derivation, which relies on the binary collision assumption.  These two derivations yield different microscopic expressions for the stress tensor, which we compare using simulations.  This provides a quantitative bound on the regime where binary collisions dominate momentum transport and reveals that most realistic granular flows occur in the region of phase space where the binary collision assumption does not apply.
\end{abstract}
\maketitle

Granular materials have emerged as an ideal non-equilibrium system, offering the opportunity to test statistical physics in settings where the underlying assumptions can be closely examined~\cite{reviewkad,reviews, reviews1}.  A central objective for granular flows is the formation of a hydrodynamic description, which is essential for both theories and applications.  To achieve this goal, there has been considerable interest in describing granular flows using extensions of the kinetic theory of dilute gases~\cite{ferzigerbook}.  This approach has successfully predicted important characteristics of dilute granular flows in situations where grain-grain interactions are dissipative and thermal effects are negligible~\cite{kinetictheories, kinetictheories1, lun, jenkins, duftypre, lutskopre}.

Of particular interest is the prediction of Bagnold's scaling, which states that the stress tensor scales with the square of the shear rate~\cite{bagnold}.  Although kinetic theory should only hold for dilute systems, experimental~\cite{pouliquen} and numerical~\cite{chevoir1a, silberts, silberts1} observations of Bagnold's scaling in dense systems have fostered hopes that kinetic theory can be applied beyond our naive expectations~\cite{mitarai} .  However, kinetic theory depends on the assumption that rigid grains, or ``hard-spheres'', interact only through instantaneous binary collisions~\cite{vannoijebook, breystat}, an assumption that will eventually break down in dense systems.  

Recently, it has become clear~\cite{lemaitre,gaj1,dacruz} that Bagnold's scaling holds on quite general grounds, which are not limited by the assumptions of kinetic theory and do not require the binary collision assumption.  Likewise, the structure of the hydrodynamic equations is largely set by conservation laws, and not by any of the assumptions that are specific to kinetic theory.  These observations, though elementary, suggest that a direct comparison of macroscopic properties with the predictions of kinetic theory is not a reliable validation tool.  Instead, tests of kinetic theory must rely on a detailed analysis of its underlying assumptions, and tailored measurements at the microscopic level.

In this paper we use simulations to quantitatively test when the binary collision assumption is appropriate to describe momentum transport in hard-sphere granular flows.  We begin by describing how momentum transport and Bagnold's scaling can be derived from macroscopic considerations, without using kinetic theory or the binary collision assumption.  Then we present the formal derivation of momentum transport using kinetic theory.  This comparison reveals that the microscopic formula for the stress tensor differs in each approach.  We measure the ratio of the pressures predicted in each derivation, thereby testing a necessary condition for the validity of kinetic theory.

\section{Momentum Transport without Kinetic Theory}
\label{woutkinetic}

A primary objective of statistical mechanics is to provide 
microscopic or molecular justifications for macroscopic equations.
At the hydrodynamic level we are interested in the macroscopic evolution of densities of the
extensive variables mass, momentum and energy.

In granular materials, mass and momentum are conserved, but energy
is lost in each collision. An equation of motion for energy can still be written
and involves a dissipative term.  We will not focus on the energy equation in this discussion,
but include it below for completeness.

Before providing microscopic forms of the conservation equations,
we recall the continuum forms.
Mass conservation reads:
\begin{equation}
\label{eqn:hydro:mass}
\frac{\partial \rho}{\partial t} = -\frac{\partial}{\partial r^\alpha} \left( \rho V^\alpha \right),
\end{equation}
where Greek superscripts represent components and repeated indices are summed.  The mass density is denoted as $\rho({\bf r},t)$ and the fluid streaming velocity has components $V^\alpha({\bf r},t)$. 
These quantities are functions of the position ${\bf r}$ and time $t$.
Similarly, conservation of momentum reads:
\begin{equation}
\frac{\partial }{\partial t } \left( \rho V^\alpha \right) =  -\frac{\partial}{\partial r^\beta} \left[ \rho V^\alpha V^\beta + \Sigma^{\alpha \beta} \right], 
\label{consmomentum}
\end{equation}
which defines the stress tensor $\Sigma^{\alpha\beta}({\bf r},t)$.

Energy is not conserved in granular materials.  Therefore, it is not possible to rely
only on the conservation of energy to define hydrodynamic equations.
However, it is expected that the equation for the energy can be written as
a ``conservative'' part plus a dissipative part. It thus takes the form:
\begin{equation}
\label{eqn:hydro:energy}
\frac{\partial}{\partial t} \left( \rho \, e \right)
=\frac{\partial}{\partial r^\alpha} \left[ \rho \,e \,V^\alpha
+J_Q^\alpha +\Sigma^{\alpha \beta} V^{\beta} \right]
+\zeta,
\end{equation}
where $e$ is the energy density, $J_Q^\alpha$ denotes the heat flux, and $\zeta$ is the dissipative term.

\subsection{The Microscopic Connection}
Connecting the macroscopic fields in the hydrodynamic description to the microscopic 
motion of individual grains in a microcanonical formulation is non-trivial and has been the focus of recent studies~\cite{Goldhirsch, Goldhirsch2}.
A microcanonical form of conservation equations begins with the introduction of a coarse-graining function ${\mathcal G}$, which allows us to define a continuous density functional:
\begin{equation}
\rho({\bf r},t) = \sum_i m_i \, {\mathcal G}({\bf r}-{\bf r_i}),
\label{eqn:rho}
\end{equation}
where $m_i$ and ${\bf r_i}$ are the mass and position of grain $i$, and the sum is over all grains in the material.
At the microscopic level, mass transport corresponds to the equation of motion for $\rho$.  Taking a time derivative in Equation~(\ref{eqn:rho}) and performing elementary algebra leads to
\begin{equation}
\frac{\partial \rho}{\partial t} = -\frac{\partial}{\partial r^\alpha} \sum_i m_i v_i^\alpha {\mathcal G}({\bf r}-{\bf r_i})
\end{equation}
where $v_i$ is the velocity of grain $i$.  This equation, combined with Equation~(\ref{eqn:hydro:mass}) defines the momentum density 
\begin{equation}
\rho({\bf r},t) V^\alpha({\bf r},t) = \sum_{i=0}^{N} m_i v_i^{\alpha} \mathcal{G}({\bf r} - {\bf r_i}).
\label{momdensity}
\end{equation}

In an analogous way, the hydrodynamic equation for momentum flux can be determined by taking the time derivative of the momentum density.
The exact microscopic expression for the stress tensor is then determined by rewriting the resulting expression in terms of a divergence.  This process is strictly algebraic and does not require any assumptions regarding the inter-particle forces~\cite{Goldhirsch}.  It yields:
\begin{equation}
\label{stressflux}
\Sigma^{\alpha \beta}({\bf r},t) = \sum_{i=0}^N  \mathcal{G}({\bf r} - {\bf r_i})m_i (v^\alpha_i-V^\alpha) ( v^\beta_i-V^\beta) + \frac{1}{2} \sum_{\{i,j\}=0}^{N_c}  \sigma_{ij}^\alpha F_{ij}^\beta \int_0^1 \mathcal{G}({\bf r} - {\bf r_i} + s {\bf \sigma}_{ij})\, ds ,
\end{equation}   
where the sum is over all $N_c$ contacts between grains, ${\bf \sigma_{ij}} = {\bf r_i}-{\bf r_j}$, and $F_{ij}^\alpha$ is the contact force between grains $i$ and $j$.  

The first term in Equation~(\ref{stressflux}) represents how the velocity fluctuations of individual particles creates stress, and the second term gives the contribution from inter-grain forces.  We will be most interested in the second term, which is often called the static contribution to the stress tensor and is denoted by
\begin{equation}
\Sigma_{\mathrm{s}}^{\alpha \beta}({\bf r},t) = \frac{1}{2} \sum_{\{i,j\}=1}^{N_c}  \sigma_{ij}^\alpha F_{ij}^\beta \int_0^1 \mathcal{G}({\bf r} - {\bf r_i} + s {\bf \sigma}_{ij})\, ds.
\label{sstress}
\end{equation}
The spatial averaging of this macroscopic variable must be conducted in a different way than the spatial averaging of momentum density-- this is due to the algebraic manipulations which converted Equation~(\ref{consmomentum}) into the divergence of a tensor.

\subsection{The Origin of Bagnold's Scaling}
Equations~(\ref{momdensity})~and~(\ref{stressflux}), combined with Equation~(\ref{consmomentum}), comprise the {\em exact} relations for transport of momentum in granular materials (and any other particle or molecular materials).  The microscopic relations depend on the properties of every particle in the material and do not require any assumptions from kinetic theory.  

In most situations, it is impractical to determine the macroscopic variables from Equations~(\ref{momdensity})~and~(\ref{stressflux}), since the precise motion of each particle is not known.  
In this case constitutive relations are postulated that relate the stress tensor to the velocity field.  These constitutive relations can be tested using simulations, where the properties of each grain are known.  Once a constitutive relation for the stress tensor is determined, the conservation equation can be solved for the velocity field.

A granular simple shear flow can be characterized by the requirements that
${\partial V_x }/{\partial y} = \dot\gamma$,
where $\dot\gamma$ is the shear rate, and ${\bf V} = V_x \hat{x}$ .  
We limit the discussion here to properties of perfectly rigid grains, which corresponds
to the limit where the time-scale introduced by the shear rate $\dot\gamma^{-1}$ is much larger than the transport time of elastic waves.  This hard-sphere limit has been the primary focus for kinetic theories of granular flow~\cite{kinetictheories, kinetictheories1}.

In the case of perfectly rigid grains, the velocity dependence of the constitutive relation for the stress tensor can be determined~\cite{gaj1}.  Since the only time-scale is provided by the shear rate, the system obeys a strict invariance which applies to rigid grains in all flow regimes:  doubling the shear rate does not alter the trajectory in configuration space, the system simply proceeds twice as fast.  Therefore, the first term in Equation~(\ref{stressflux}) for the stress tensor must scale as the square of $\dot\gamma$.   
In addition, upon increasing the shear rate, the inter-grain forces will scale as the square of the shear rate, since they are related to grain accelerations.  Therefore the static stress from Equation~(\ref{sstress}) must also depend on the square of the shear rate.    

This non-Newtonian behavior of the stress tensor has been observed in both experiments~\cite{bagnold} and simulations~\cite{chevoir1a,silberts, silberts1, dacruz} of realistic granular flows, and is referred to as ``Bagnold's scaling''.  Simulations of granular materials in simplified geometries~\cite{campbell} also suggest that most realistic granular flows occur in the rigid grain regime where, through the time invariance, Bagnold's scaling {\em must hold}.  Therefore, considering the properties of perfectly rigid grains gives insight to the dynamics of realistic granular flows.

\section{Momentum Transport with Kinetic Theory}
\label{wkinetic}

The kinetic theory of frictionless, hard-sphere granular materials leads to a continuum hydrodynamic description, beginning with assumptions about the microscopic interaction between grains~\cite{lun, jenkins, duftypre, lutskopre}.  Since mass and momentum are conserved, the transport equations are the same as in Equations~(\ref{eqn:hydro:mass})~and~(\ref{consmomentum}).  However, because kinetic theory is based on certain microscopic assumptions, the value of the stress tensor in Equation~(\ref{consmomentum}) is determined within these assumptions.       

Recently, a formal construction of the transport equations in the hard-sphere kinetic theory framework has emerged~\cite{duftypre, lutskopre, vannoijebook, breystat}.  This derivation is an extension of the kinetic theory of classical gases (or dense gases) where energy is conserved.  The novel aspect is the introduction of a restitution coefficient $\alpha$, which approximates the energy dissipation in individual collisions. 
The first step in the construction assumes that the only interactions between grains consist of instantaneous binary collisions~\cite{vannoijebook, breystat}.  In each binary collision, momentum is conserved and energy is dissipated via the restitution coefficient $\alpha$:
\begin{equation}
({\bf v_2} - {\bf v_1}) \cdot \hat\sigma_{12} = -\alpha ({\bf v'_2} - {\bf v'_1}) \cdot \hat\sigma_{12},
\label{binarycollision}
\end{equation}
where $\{ {\bf v'_1}, {\bf v'_2} \}$ denote pre-collisional velocities and $\{ {\bf v_1}, {\bf v_2} \}$ denote post-collisional velocities of the colliding grains indexed by $1$ and $2$.  The normal unit vector $\hat\sigma_{12} = \frac{{\bf r_1} - {\bf r_2}}{|{\bf r_1}-{\bf r_2}|}$ limits the dissipation to occur in the normal components of velocity.  By enforcing Newton's equation in each collision, the dynamical rule from Equation~(\ref{binarycollision}) determines a collisional impulse
\begin{equation}
{\bf I}_{\mathrm{bc}} = (1+\alpha) \mu_{12} \left[ ({\bf v'_1} - {\bf v'_2}) \cdot \hat{\sigma}_{12} \right] \hat{\sigma}_{12},
\label{binaryimpulse}
\end{equation}
where $\mu_{12}$ is the reduced mass.
This impulse acts equally and oppositely on each grain involved in the collision.

\subsection{Pseudo-Liouville Equation and the BBGKY hierarchy}
After the microscopic interactions have been specified, the formal derivation of hard sphere kinetic theory relies on the pseudo-Liouville equation~\cite{vannoijebook, breystat}.  This equation determines the time dependence of the $N$-particle probability distribution function (pdf) $f^{(N)}$,  where $f^{(N)}$ is defined such that $f^{(N)} d{\bf v_j} d{\bf r_j}$ is the probability that the particle $j \in [1,N]$ has a position ${\bf r_j}$ and velocity ${\bf v_j}$.
The pseudo-Liouville equation can be derived by directly considering the dynamical rule in Equation~(\ref{binarycollision}) and using the assumption that only binary collisions occur.  The result is 
\begin{equation}
(\frac{\partial}{\partial t} + \mathcal{L}) f^{(N)} = 0,
\label{pseudoliouville}
\end{equation} 
where $\mathcal{L}$ is a functional of $f^{(N)}$.  The exact form for $\mathcal{L}$ is not important in this paper, and can be found in previous works~\cite{vannoijebook, breystat}.

Once the pseudo-Liouville equation is derived, the time dependence of the $s$-particle pdf $f^{(s)}$ can be determined by integration.  The function $f^{(s)}$ is determined directly from $f^{(N)}$ by the relation
\begin{equation}
f^{(s)} = \frac{N!}{(N-s)!}\int \prod_{i=s+1}^N d{\bf v_i} d{\bf r_i} f^{(N)},
\end{equation}
and by integrating Equation~(\ref{pseudoliouville}) over variables $\prod_{i=s+1}^N d{\bf v_i} d{\bf r_i} $.  The pseudo-Liouville equation yields the time dependence of $f^{(s)}$ as a function of an integral over $\mathcal{L}$, which can be determined without further assumption~\cite{breystat}.  

The system of $N$ equations for the $s$-particle pdfs that results from integrating the pseudo-Liouville equation is called the BBGKY hierarchy, and the most important of these equations for momentum transport in granular materials is the pdf for $s=1$.  This forms the basis for a derivation of transport equations and reads~\cite{breystat, duftypre, lutskopre}
\begin{equation}
\left(\frac{\partial}{\partial t} + v_1^\alpha \frac{\partial}{\partial r^\alpha} \right) f^{(1)}({\bf r_1},{\bf v_1},t) = T\left( {\bf r_1},{\bf v_1} \right),
\label{bbgkyone}
\end{equation}
where $T$ is an operator that describes how the one-particle pdf changes due to interactions between grains.  When the binary collision assumption is made, this operator only depends on the two-particle pdf.  The functional form is
\begin{eqnarray}
T\left( {\bf r_1},{\bf v_1} \right) &=& \sigma^2 \int d{\bf v_2} \int d\hat{\sigma} \Theta(\hat{\sigma} \cdot {\bf g}) (\hat{\sigma} \cdot {\bf g}) \nonumber \\
  &\times& [ \alpha^{-2} f^{(2)}\left({\bf r_1},{\bf r_1}-{\bf \sigma},{\bf v'_1},{\bf v'_2},t\right) 
     - f^{(2)}\left({\bf r_1},{\bf r_1}+{\bf \sigma},{\bf v_1},{\bf v_2},t\right) ],
\label{Teqn}
\end{eqnarray}
where ${\bf g} = {\bf v_1}-{\bf v_2}$ and ${\bf \sigma} = \sigma \hat{\sigma}$ is the continuous variable representing the displacement vector between two grains in contact.  

The collision operator $T$ is often called the Enskog collision operator, since Enskog originally derived it for classical gases ($\alpha=1$) using a less formal approach~\cite{ferzigerbook}.  The collision operator encodes the change in the one-particle pdf due to collisions that produce a particle with velocity ${\bf v_1}$ (first term) and remove a particle with velocity ${\bf v_1}$ (second term).  The step function $\Theta$ limits the integral to pairs of colliding grains, and the factor of $(\hat{\sigma} \cdot {\bf g})$ sets the rate of collisions.     

The derivation of $T$ explicitly relies on the binary collision assumption.  If this assumption is not made, additional factors must be included in $T$, which are proportional to higher $s$-particle pdfs.  For example, if we assume that up to $n$ grains can interact simultaneously, then terms proportional to each of the $s$-particle pdfs, with $s \leq n$, must be included.  Obviously, including many more terms will make the mathematics much more difficult, and values of $n$ larger than two have not been considered in detail.

\subsection{Determining the stress tensor}
Kinetic theory furnishes Equations~(\ref{bbgkyone}) and (\ref{Teqn}) that describe the time dependence of the one-particle distribution function through an evaluation of the microscopic interactions between grains.  The only assumption is that the interactions consist solely of binary collisions.  The next step is to connect the microscopic formulation to macroscopic transport equations, in particular momentum transport.  

Using the convention that $\rho({\bf r},t) = \int m f^{(1)}\left({\bf r},{\bf v},t \right)d{\bf v} $, where $m$ is the average grain mass, the momentum density is related to the one-particle pdf through the relation ${\bf \rho}({\bf r},t) {\bf V}({\bf r},t) = \int m {\bf v} f^{(1)}\left({\bf r},{\bf v},t\right)d{\bf v} $.  Therefore, by multiplying each side of Equation~(\ref{bbgkyone}) by $m_1 {\bf v_1}$, and then integrating over ${\bf v_1}$, we obtain an equation for the time dependence of the momentum density.  This also results in an equation for the microscopic form of the stress tensor, as predicted by kinetic theory~\cite{duftypre, lutskopre}
\begin{eqnarray}
\label{ktstress}
\Sigma^{\alpha \beta}({\bf r},t) &=& \int d{\bf v} m ( v^\alpha -  V^\alpha)( v^\beta -  V^\beta) f^{(1)}({\bf r},{\bf v}) \\ + \frac{1+\alpha}{4} &m& \sigma^3 \int d{\bf v_1} d{\bf v_2} d\hat{\sigma} \Theta(\hat{\sigma} \cdot {\bf g}) (\hat{\sigma} \cdot {\bf g})^2 \hat{\sigma}^\alpha \hat{\sigma}^\beta \int_0^1 ds f^{(2)}[{\bf r}-(1-s){\bf \sigma},{\bf r}+s{\bf \sigma},{\bf v_1},{\bf v_2}]. \nonumber
\end{eqnarray}
As in the more general case (Equation~(\ref{stressflux})), kinetic theory predicts that the stress consists of two contributions:  one from the velocity fluctuations of individual grains and one from contacts between grains (which we earlier called the static stress).    

In the case of kinetic theory, the prediction for the static stress only includes contributions from binary collisions.  As we defined static stress based on Equation~(\ref{stressflux}), we refer to the second term in Equation~(\ref{ktstress}) as the ``collisional'' stress, and denote it as   
\begin{equation}
\Sigma_{\mathrm{bc}}^{\alpha \beta}({\bf r},t) = \frac{1+\alpha}{4} m \sigma^3 \int d{\bf v_1} d{\bf v_2} d\hat{\sigma} \Theta(\hat{\sigma} \cdot {\bf g}) (\hat{\sigma} \cdot {\bf g})^2 \hat{\sigma}^\alpha \hat{\sigma}^\beta \int_0^1 ds f^{(2)}[{\bf r}-(1-s){\bf \sigma},{\bf r}+s{\bf \sigma},{\bf v_1},{\bf v_2}].
\label{bcstress}
\end{equation}
The subscript ``bc'' reminds us that this is the static stress predicted by kinetic theory, where only binary collisions are considered.

Equation~(\ref{ktstress}) also reveals that Bagnold's scaling holds for the stress tensor derived from kinetic theory.  Each of the terms scales with two factors of a velocity, and the only velocity scale is given by the shear rate.  Thus the stress tensor must be proportional to the square of the shear rate.  This is not surprising, since any shear flow of rigid grains (whether or not the assumptions of kinetic theory apply) will obey Bagnold's scaling.  Therefore, although kinetic theory is consistent with Bagnold's scaling, we must remember that the observation of Bagnold's scaling in a granular flow does not imply that kinetic theory holds.

\section{Numerical Tests}
\label{tests}
The form of the momentum transport equation depends only on the conservation of momentum and therefore holds whether or not kinetic theory is used.  However, the definition of the stress tensor is dependent on the microscopic assumptions made in kinetic theory.  Using a kinetic theory approach we would predict Equation~(\ref{ktstress}), and using a general approach we would predict Equation~(\ref{stressflux}).  In this section we test whether the stress tensor predicted from kinetic theory gives the same numerical value as that predicted by the general approach.  This is a necessary condition for kinetic theory to make accurate predictions.

To make the comparison we simulate a 2D polydisperse simple-shear flow of rigid grains using the Contact Dynamics (CD) algorithm~\cite{cdalgo, cdalgo1}.  The polydispersity of the grains is characterized by an equal probability to have a grain diameter within the range $\sigma \pm \Delta$, with $\Delta/\sigma = 0.26$.  The motion of each grain is set by Newton's equations, with the forces between grains determined by enforcing the collision rule in Equation~(\ref{binarycollision}) for each pair of contacting grains.  A set timestep $dt=10^{-5}$ is used, and all collisions that occur in the timestep are considered simultaneously.  Shear flow is produced using Lees-Edwards boundary conditions along with the SLLOD equations of motion.    More details on our specific algorithm can be found in~\cite{gaj1}, where we also observe that (i) Bagnold's scaling always holds, (ii) the polydispersity restricts crystallization, and (iii) the Lees-Edwards boundary conditions guarantee translation invariance.

The CD algorithm is the natural choice for this particular numerical test.  It guarantees that the grains are treated as perfectly rigid, so we need not worry about the value of the stiffness, as in a soft-sphere simulation~\cite{campbell, simulations, softspheres}.  Additionally, because the CD algorithm uses a constant timestep, it can simulate perfectly rigid granular flows at any density (below random close packing) and is not limited by the assumption that only binary collisions occur, as is the case for event driven simulations~\cite{simulations, eventdriven}.

\subsection{Comparing stress tensors}

To compare the stress tensors, we determine averages over the entire simulation cell.  This simplifies the analysis since the averaging function is then $\mathcal{G}=A^{-1}$, where $A$ is the area of the simulation cell.  Additionally, because the simulations are translationally invariant, the positional dependence of the one and two-particle pdfs just provides a factor of $A^{-1}$.  

As a result, we see that the first term in Equations~(\ref{stressflux})~and~(\ref{ktstress}) are identical.  This is because the integral $\int d{\bf v} f^{(1)}({\bf r},{\bf v}) = \int d{\bf v} A^{-1} f({\bf v})$ plays the role of averaging over all grains, and can be replaced by a sum over grains.  This reproduces the first term in Equation~(\ref{stressflux}). 

Using the same simplifications, the second terms in the stress equations yield
\begin{equation}
\Sigma_{\mathrm{s}}^{\alpha \beta} = \frac{1}{2 A} \sum_{\{i,j\}=1}^{N_c}  \sigma_{ij}^\alpha F_{ij}^\beta,
\label{firstsimpsstress}
\end{equation}
and
\begin{equation}
\Sigma_{\mathrm{bc}}^{\alpha \beta} = \frac{1}{2 A} \sigma^2 \int d{\bf v_1} d{\bf v_2} d\hat{\sigma} \Theta(\hat{\sigma} \cdot {\bf g}) (\hat{\sigma} \cdot {\bf g}) f^{(2)}[{\bf v_1},{\bf v_2}] \sigma^\alpha I_{\mathrm{bc}}^\beta ,
\label{firstsimpbcstress}
\end{equation}
where we have inserted the binary impulse from Equation~(\ref{binaryimpulse}) into the collisional stress, with the reduced mass replaced by one-half the average mass.  Equation~(\ref{firstsimpsstress}) is the total static stress, and the Equation~(\ref{firstsimpbcstress}) is the static stress predicted by kinetic theory using the binary collision assumption (the collisional stress).  Realizing that $\sigma^2 \int d{\bf v_1} d{\bf v_2} d\hat{\sigma} \Theta(\hat{\sigma} \cdot {\bf g}) (\hat{\sigma} \cdot {\bf g}) f^{(2)}$ is the collision rate~\cite{collisionrate}, then the collisional stress can be written as
\begin{equation}
\Sigma_{\mathrm{bc}}^{\alpha \beta} = \frac{N_c}{dt} \times \langle \frac{\sigma^\alpha I_{\mathrm{bc}}^\beta}{2 A} \rangle,
\end{equation}
where the first term accounts for the collision rate (the number of collisions $N_c$ per simulation timestep $dt$) and the second term gives the average value of the impulse in each collision, as determined in Equation~(\ref{binaryimpulse}).  This can be transformed to a sum over collisions written as 
\begin{equation}
\Sigma_{\mathrm{bc}}^{\alpha \beta} = \frac{1}{2 A} \sum_{\{i,j\}=1}^{N_c} \sigma^\alpha_{ij} \left(F_{\mathrm{bc}}\right)^\beta_{ij},
\label{finalbcstress}
\end{equation}
where ${\bf F}_{\mathrm{bc}} = {\bf I}_{\mathrm{bc}}/dt$ is the total force exerted through binary collisions in a single timestep $dt$ of the simulations.

With the derivation of Equations~(\ref{firstsimpsstress})~and~(\ref{finalbcstress}), we have succeeded in re-writing the static and collisional stress in forms that can be easily computed in simulations.  We see that the form of the collisional stress is exactly the same as the static stress, {\em with the total contact force replaced by the binary collision force}.  This is because the only assumption that kinetic theory makes to derive the stress tensor in Equation~(\ref{ktstress}) is the binary collisions assumption.  

\begin{figure}
\begin{center}
\psfrag{tl}{\Huge{$\langle F \rangle/\langle F_{\mathrm{bc}} \rangle$}}
\scalebox{0.45}{\includegraphics{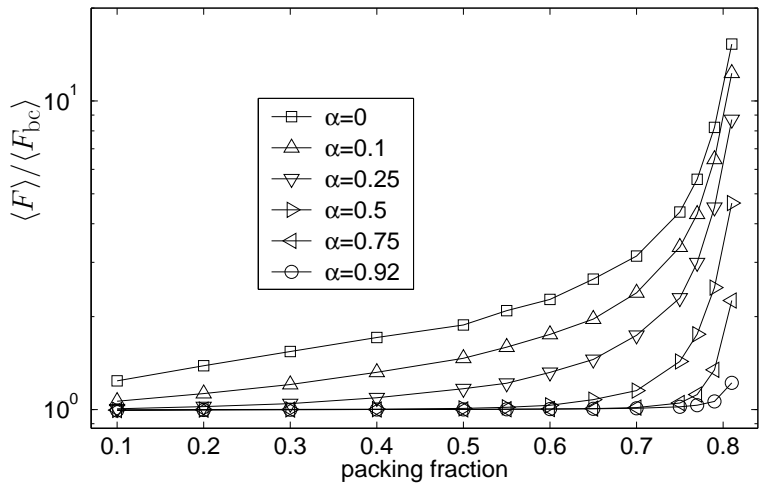}}
\caption{\label{ratio} The ratio of average contact force to the average binary contribution.  Values equal to one correspond to restitution coefficients $\alpha$ and packing fraction $\nu$ where kinetic theory is able to make accurate predictions.}
\end{center}
\end{figure}

Therefore, in order to test the accuracy of kinetic theory, we simply need to measure the magnitude of the average contact force ${\bf F}$ and compare it to the average magnitude of the collisional force ${\bf F}_{\mathrm{bc}}$.  A necessary test of the validity of kinetic theory is that this ratio is equal to one.  In Figure~\ref{ratio} we present data for the ratio $\langle F\rangle /\langle F_{\mathrm{bc}} \rangle$ for all values of the restitution coefficient and packing fraction we have investigated.  The packing fraction $\nu$ is equal to the area of grains divided by the total simulation area.  The average was taken over $2000$ timesteps in steady state.
The different symbols correspond to different resitution coefficents, with the higher curves corresponding to decreasing $\alpha$ (i.e. increasing dissipation) and therefore increasing inaccuracy of predictions based on kinetic theory.

We see from Figure~\ref{ratio} that for many combinations $\{\alpha,\nu\}$, the ratio of forces is equal to one and kinetic theory is applicable.  However, for larger values of $\nu$ and smaller values of $\alpha$, the ratio becomes larger than one and kinetic theory can not be used to predict the stress tensor.  Notice that the failure of kinetic theory does not depend on any approximation scheme used to calculate constitutive relations for the stress tensor, rather it is fundamentally due to the assumption of binary collisions, which is made at the beginning of the formal derivation of kinetic theory.

One of the more surprising conclusions of the data in Figure~\ref{ratio} is that the validity of kinetic theory depends on the pair of variables $\{\alpha,\nu\}$ and not just the density.  For flows at high density, a high restitution coefficient tends to keep grains well spaced so that only binary collisions occur; for flows at low restitution, collisions tend to cluster grains that have previously interacted, causing kinetic theory to fail even at low densities.  

\begin{figure}
\begin{center}
\psfrag{1-\alpha}{\Huge{$1-\alpha$}}
\psfrag{\nu}{\Huge{$\nu$}}
\scalebox{0.45}{\includegraphics{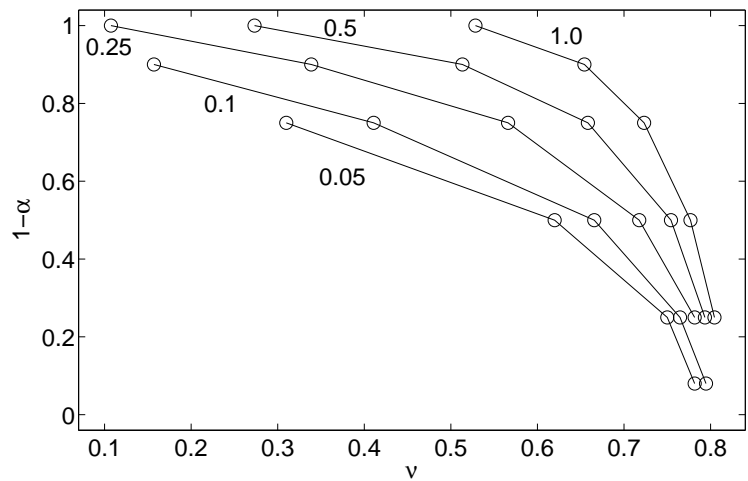}}
\caption{\label{contourplot} Contours of constant $\langle F \rangle/\langle F_{\mathrm{bc}} \rangle - 1$, with the value written beside each curve, for all restitution coefficients $\alpha$ and packing fractions $\nu$ we have investigated.  The different contours correspond to values of $\alpha$ and $\nu$ where error induced by the binary collision assumption is $5\%$ ($0.05$), $10\%$ ($0.1$), $25\%$ ($0.25$), $50\%$ ($0.5$), and $100\%$ ($1$).}
\end{center}
\end{figure}

Kinetic theory is only useful when it accurately predicts transport of momentum.  In Figure~\ref{contourplot} we plot contours of $\langle F \rangle/\langle F_{\mathrm{bc}} \rangle - 1$ in the phase space of $1-\alpha$ and $\nu$.  A plot of this quantity, which is equal to zero if only binary collisions occur, allows us to easily quantify the error induced by the binary collision assumption.  The first thing to notice about the phase plot is that there is a wide range of restitution and density where the error induced by the binary collision assumption is less than $5 \%$-- this corresponds to the area below the curve labeled $0.05$.  However, depending on the required numerical accuracy, the binary collision assumption begins to make poor predictions at large density.  Since most natural flows occur above a packing fraction of $0.75$, the applicability of kinetic theory to natural flows is quite limited.  

The numerical results in Figure~\ref{contourplot} provide a quantitative test of a necessary condition for kinetic theory to be applied to granular flows.  It is necessary that the collisional stress tensor, which is predicted by kinetic theory where only binary collisions are assumed to hold, be equal to the actual static stress tensor.  This is only achieved for a limited amount of phase space.  In the other regions, high dissipation and high density causes clusters of interaction grains to form, thereby limiting a kinetic theory approach.

\section{Conclusion}
We have presented a general derivation of the momentum transfer equation and compared it to the kinetic theory derivation for rigid granular flows.  The form of the transfer equation is based on conservation of momentum and therefore the same in both derivations.  However, the predictions for the value of the stress tensor varies.  Kinetic theory is accurate when the two stress tensors are equal, which occurs when the average contact force is equal to the average collisional force and interactions between grains are strictly binary.  We have quantitatively identified the range where kinetic theory is invalidated by considering a ratio of the total and collisional contact forces, and we find that kinetic theory is unable to make accurate predictions for a certain range of density and restitution coefficient.  

In the regions where kinetic theory fails, clusters of contacting grains form that interact through force chains.  The presence of force chains becomes the major contributor to the contact forces between grains, and these force chains must be incorporated in theory to properly characterize momentum transport in granular materials.

This work was supported by the William. M. Keck Foundation, the MRSEC program of the NSF under Award No. DMR00-80034
, the James. S. McDonnell Foundation, NSF Grant No. DMR-9813752, and the David and Lucile Packard Foundation.

\end{document}